# Adaptive cross-country optimisation strategies in thermal soaring birds

**Authors:** Göksel Keskin[1,2], Olivier Duriez[3], Pedro Lacerda[1,2], Andrea Flack[4], Máté Nagy[1,2,5]

**Affiliations:**

[1]MTA-ELTE Lendület Collective Behaviour Research Group, Hungarian Academy of Sciences, Budapest, Hungary
[2]Department of Biological Physics, Eötvös Loránd University, Budapest, Hungary
[3]CEFE, Univ Montpellier, CNRS, EPHE, IRD, Montpellier, France
[4]Collective Migration Group, Max-Planck Institute of Animal Behavior, Konstanz, Germany
[5]Max-Planck Institute of Animal Behavior, Konstanz, Germany

# Abstract

Thermal soaring enables birds to perform cost-efficient flights during foraging or migration trips. Yet, although all soaring birds exploit vertical winds effectively, this group contains species that vary strongly in their morphologies. Aerodynamic rules dictate the costs and benefits of flight, but, depending on their ecological needs, species may use different behavioural strategies. To quantify these morphology-related differences in behavioural cross-country strategies, we compiled and analysed a large dataset, which includes data from over a hundred individuals from 12 soaring species recorded with high frequency tracking devices. We quantified the performance during thermalling and gliding flights, and the overall cross-country behaviour that is the combination of both. Our results confirmed aerodynamic theory across the 12 species; species with higher wing loading typically flew faster, and consequently turned on a larger radius, than lighter ones. Furthermore, the combination of circling radius and minimum sink speed determines the maximum benefits soaring birds can obtain from thermals. Also, we observed a spectrum of strategies regarding the adaptivity to thermal strength and uncovered a universal rule for cross-country strategies for all analysed species. Finally, our newly described behavioural rules can provide inspirations for technical applications, like the development of autopilot systems for autonomous robotic gliders.

# Main

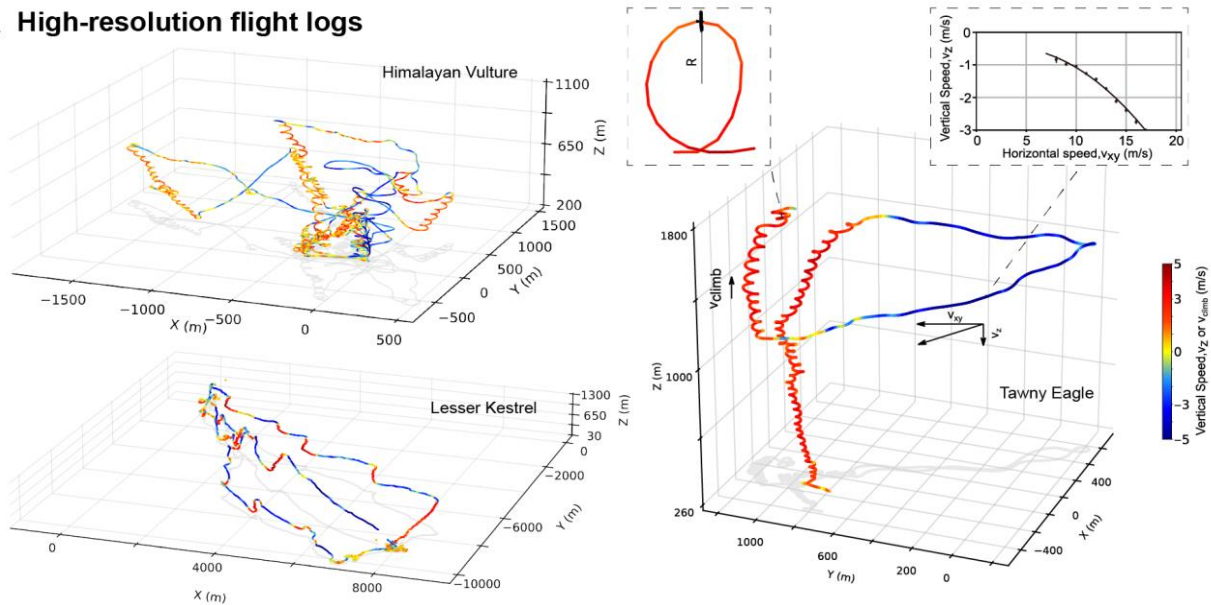

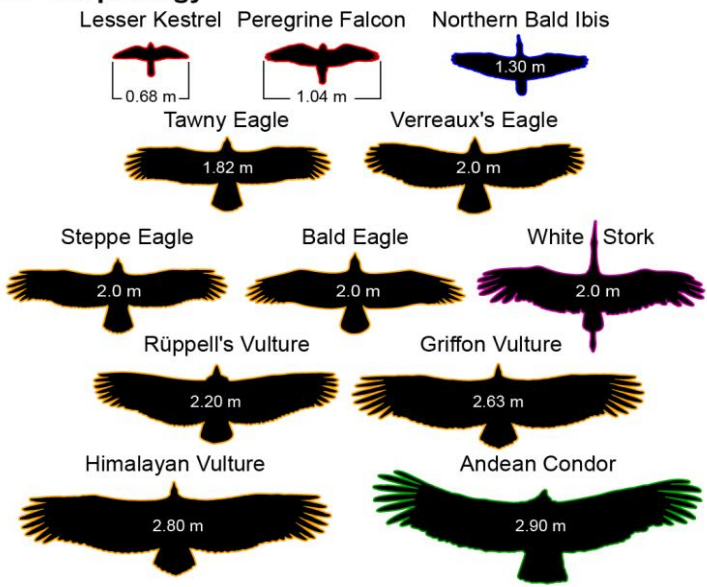

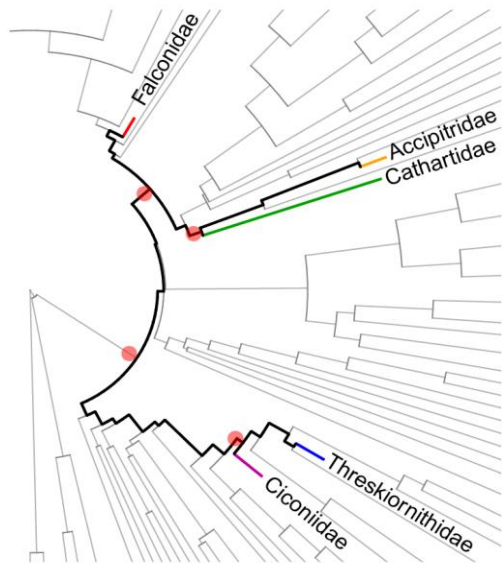

**Figure 1: Overview of our study.** High-frequency GPS data sets of 12 bird species were collected to understand unpowered flight mechanics and behaviour (A) and their connections to morphological traits (B) and phylogenetic relatedness (C). **A)** Flight logs (from 0.3 Hz to 10 Hz; examples shown for 3 individuals from different species) were used to create glide polar (using horizontal and vertical speed components of gliding flight), to obtain flight characteristics in thermals (climb speed and radius of circles), and to understand the variation in flight strategy according to the daily thermal strength. **B)** Visualization of the wing shapes of all species in our study showcasing a large variation in wingspan (depicted on the pictures). Outline colours indicate the taxonomic group the species belongs to (shown on C). **C)** Phylogenetic tree of bird species [1] marked with different colours for the families of species studied. Red circles show last common ancestors for different families, and the network distance and the phylogenetic closeness were used to comprehend the differences in soaring optimization strategies. (See also Table S1.)

Flying is energetically costly but many species have adapted their morphology and behaviour to cope with the different requirements of this aerial lifestyle. Large, heavy bird species have developed energy-efficient flight modes, such as soaring-gliding flight [2,3] because the energetic costs of flapping wings increase with body mass [4–6]. Although the aerodynamic theory behind soaring is well established, soaring birds exhibit a strong ecological, behavioural, and morphological diversity that is usually neglected in research but most likely also affects their flight performances.

During soaring, birds gain altitude when circling in thermal convective updrafts (commonly referred to as thermals: localized regions of rising, buoyant air heated by sunlight[7,8]). Thermalling is followed by gliding flight (inter-thermal flight), where birds descend while traveling horizontal distances [9,10]. During this gliding phase, birds can adjust their gliding angle (i.e. steepness of the descending glide) which determines their horizontal (gliding) airspeed and their vertical (sinking) speed. This relationship is also known as the glide polar [11,12]. Previous studies have shown that the performance of a glider (bird or aircraft) depends on its wing shape [11,13,14]. More specifically, i) the horizontal speed that ensures the maximal horizontal travel distance from a given height depends ($V_{xy}^{Best\ Glide}$) on the wing loading, ii) the glider's lift is a function of the wing's aspect ratio [12], and iii) its turning radius during thermalling is directly proportional to wing loading [12]. Thus, all these morphological factors drive flight performance but disentangling the importance of each factor separately and for different species is challenging.

Previous studies of soaring flight used motor gliders [5,15], or radars [8,16–18] to examine the flight performance of free-flying birds. Nowadays, modern biologging techniques allow us to obtain not just high-precision measurements of the birds' positions and movements in three dimensions but enable us to accurately measure the morphological features of each bird and species [19–23]. Here we compiled a large tracking data set (detailed in Table S1), that contains detailed flight records of 12 bird species, belonging to 5 families with very different lifestyles (*Accipitridae, Cathartidae, Falconidae, Threskiornithidae, Ciconiidae*). The dataset includes scavengers (3 (old-world) vultures and 1 (new-world) condor that look for food on the ground from high altitudes;), predators (searching for mobile prey that attack in the air (1 large falcon), on the ground (3 eagles) or in water (1 sea eagle), and 3 species foraging on insects or small mammals that migrate long distances (1 small falcon, 1 stork and 1 ibis). Despite these behavioural differences, all species rely heavily on soaring flight and have broad, elongated wings (relative to their body mass) with relatively similar aspect ratios (between 6.27 and 8.46). In addition, they differ in wingspan, wing loading and body mass (by almost 2 orders of magnitude, see Table S2), likely generating interspecific variations in flight behaviour. Although our main goal is to compare different species, we also investigated individual trajectories of *Gyps fulvus*, where the large number of tracked individuals tracked allowed us to examine intraspecific variations in flight behaviour and compare them with the general species-level findings.

We compared the soaring and gliding behaviours of the different species by focusing on the performance and optimization of cross-country flights under different thermal conditions. We quantitatively characterized the species' flight performance and cross-country optimization strategies, and compared the experimental results to the theoretical expectations of Pennycuick's flight tool [12] and the MacCready theory [24]. We explored the potential use of three cross-country strategies that are optimized for different goals: (1) a strategy based on MacCready's theory that guarantees a maximal cross-country speed by adjusting gliding (horizontal) speed to thermal strength (ascending speed) according to the species' respective polar curve; (2) a strategy that maximizes travel distance from a given height using a gliding speed independent of thermal strength (i.e. choosing a horizontal speed corresponding to the best glide); or (3) a mixed strategy that combines the previous two. We predict that most species adapt their cross-country strategies according to the prevailing thermalling conditions [25,26]. A previous study, using a similar comparative approach, explored how gliding airspeed relates to the species' morphology while also testing whether this relationship also depends upon the risk of not finding a new thermal [18]. Yet, their flight data came from radar tracks only, without detailed individual morphological variations. In addition, their framework relied heavily on Pennycuick's equations [12] and could not cope with birds behaving outside of the expected optimal range (which could be very narrow in the case of weak thermals). Thus, here using high-frequency flight recordings, we tested the validity of these previous theoretical predictions from Pennycuick to then analyse cross-country strategies based on observed flight parameters (Figure 1).

## Results

### *Empirical and theoretical glide polar curves*

To determine how these foraging specializations (e.g. aerial foragers vs. large, heavy scavengers; see Table S1) and the connected morphological differences (Table S2) relate to the gliding performance of these different species, we first created effective (empirical, observed) polar curves of the 12 species using high-frequency GPS trajectories during soaring-gliding flight (Table S1 and Supplementary Dataset, Fig.S1), following a second-order approximation[25] ($f(x) = ax^2 + bx + c$). Note that Griffon vultures (*G. fulvus*) data originated from two different sources: (1) free-flying adults that were raised in captivity and trained with falconry techniques [27], and (2) free-flying wild birds of various ages [28]. The two data sets provided distinctive polar curves (mean absolute difference = 0.66 m/s, p=0.64) which is why we decided to analyse them separately.

First, we experimentally validated previous theoretical approximations by examining how similar the empirical, data-based polar curves (although those may include flapping flight as well) are to the polar curves resulting from Pennycuick's equations using the default flight tool settings for each species (i.e. default curves from Flight software, version 1.25; www.bristol.ac.uk/biology/media/pennycuick.c/flight_123.zip)[12] . We first noticed that these species-specific Pennycuick polar curves were similar between each species. To quantify the similarity, we calculated the mean absolute difference of the glide polars and found little variation between the species (mean = 0.09 m/s, standard deviation (SD) = 0.05 m/s, Fig. S2 A). Moreover, the polar curves based on Pennycuick's flight tool with default values (default curves) did not fit well the empirical data (Fig. S2 B). The mean absolute difference between the default curves and observed data points was not smaller compared to what was expected by chance when comparing to polar curves of different species ( $\langle \Delta_{default} \rangle$ = 0.98 m/s, n=12, randomization test, p = 0.5581, Fig S2 B.). Although Pennycuick also suggested that these formulas should not be used with the default parameter settings [12], many research studies only rely on the default values. Thus, we search the literature [12,29,30] for more realistic physical properties of the different species (i.e. the body drag coefficient, wing profile coefficient, and maximum lift coefficient). Using these updated parameters (shown in Table S3) the estimated glide polars ("improved" Pennycuick curves) became more diverse between the species (larger difference between the curves: mean = 0.24 m/s, SD = 0.14 m/s, Fig. S2 A.) and more similar to the observed data points ( $\langle \Delta_{improved} \rangle$ = 0.47 m/s), but still did not fit significantly better than what was expected by chance (n=12, randomization test, p = 0.5547). Yet, the data-based polar curves provided significantly better fit to the observed data points ( $\langle \Delta_{empirical} \rangle$ = 0.19 m/s, n=12, randomization test, p = 0.0488) with a large variation between the species (difference between the curves: mean = 0.58 m/s, SD = 0.27 m/s, Fig. S2 A) which is why we used those for the remaining analyses.

Two main parameters, defined by the polar curve, are crucial for understanding flight performances: (i) "minimum sink" which provides maximum gliding time from a given height (it results in a minimal rate for losing height), and (ii) the "best glide" providing maximum horizontal travel distance from a given height. The best glide is defined where the glide ratio (ratio between the horizontal and vertical speed) is the highest and can be determined by drawing a tangent from the origin (as illustrated in Figure 2). Figure 2 shows our data-based polar curves with both parameters for the 12 species, highlighting similarities between related species, although some of the similarities may come from similar morphology of unrelated species. One important aspect to notice, despite *F. naumanni* and *C. ciconia* exhibiting a similar best glide ratio (also see SI Table S4 Gliding), they flew at different airspeeds to achieve the best glide. These flight parameters allowed us to investigate the effects of morphology on flight performance. Wing loading is expected to have a crucial role both in terms of soaring and gliding efficiencies. We studied in detail the functional relationship between wing loading and glide polar characteristics (Fig 3) as previous theoretical and empirical studies of aerodynamics for birds [11,12] and aircrafts [14,31] suggest relationships. We found a positive relationship between wing loading and horizontal speed both for best glide ratio (linear regression without intercept, coefficient of determination, $R^2 = 0.96$, $p = 0.0021$, randomization test; Fig 3A) and for minimum sink speed ($R^2 = 0.86$, $p = 0.0314$; Fig 3B). Furthermore, during the soaring phase, individuals with higher wing loading typically flew faster and consequently turned with a larger radius, compared to "lighter" species. Under the assumption that bank angle (∅) is constant and equal to all birds, the relationships between circling radius and other flight parameters, such as wing loading ($R^2=0.93$, $n = 13$, $p = 0.0249$; Fig 3C), or the average horizontal speed ($R^2=0.97$, $n = 13$, $p < 0.0001$; Fig 3D) were previously reported [12,32], but here we confirm this for a much larger number of species. Besides wing loading, aspect ratio was also expected to affect flight performance during gliding, but in our set of species, the differences in aspect ratio (6.27 – 8.46) were relatively small (as compared to, for example, the variation in wing loading, 2.1-9.1 kg/m$^2$), thus not providing a large enough range to study its effect (Fig. S3).

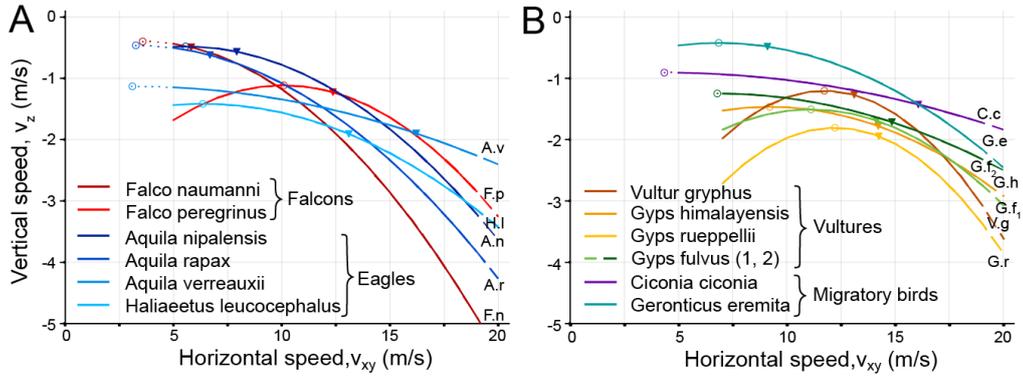

**Figure 2: Empirical polar curves fitted to the flight data presenting the most important features of the gliding.** The relationship between gliding airspeed and vertical speed for 12 species pooled into groups according to genetic relatedness and size similarity. Circles denote the minimum sink (the highest point of the parabola fitted to the entire data range). Triangles denote the "best glide" using which birds can travel the furthest from a given height. Dotted and dashed lines indicate the extrapolation of the parabola beyond the range of the used data. (See also Figures S1, S2, S4, and Table S3.)

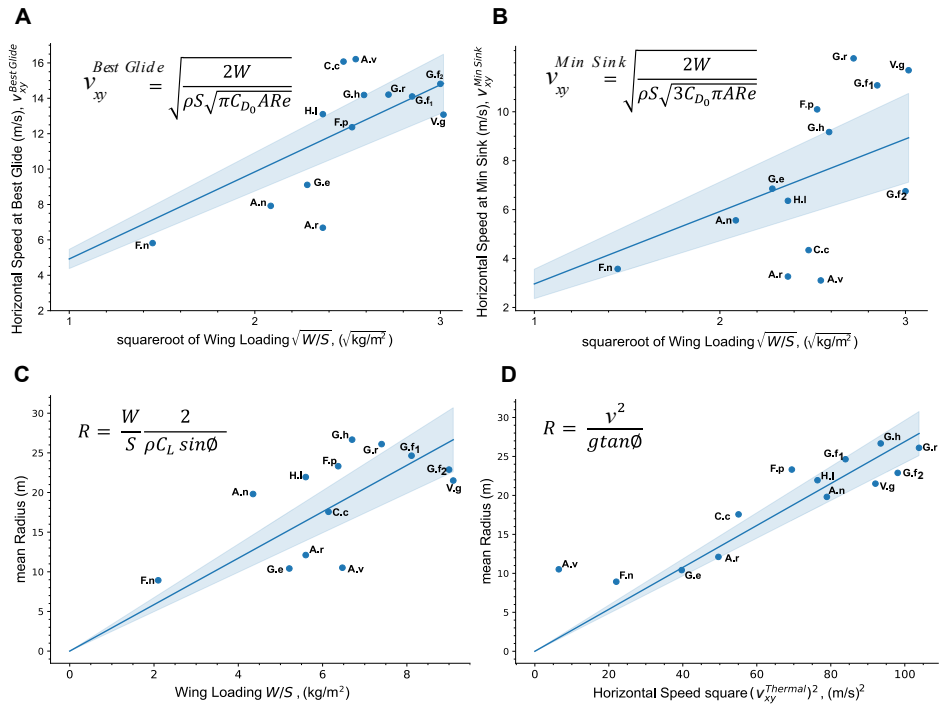

**Figure 3: Relationship between flight behaviour and morphological parameter, and comparison to theoretical predictions.** Circles present the mean values for species (with a two-letter abbreviation using the Latin name; see Table S1). Line shows linear fit to the data points with confidence bands indicated as the shaded areas. The formula presented on each plot was derived from aerodynamic theory of unpowered flight. **A-B**) Wing loading ($W/S$) determines the speed at maximum glide ratio (**A**, $R^2 = 0.96$) and minimum sink rate (vertical speed) (**B**, $R^2 = 0.88$), but wing aspect ratio ($AR$), zero-lift drag coefficient ($C_{D0}$), air density ($\rho$), and Oswald efficiency factor ($e$) also have an effect. **C-D**) Assuming the bank angle ($\emptyset$) is constant and equal to all birds (although it may be a strong assumption[33]), the radius of the turn in steady flight, is related to wing loading ($W/S$, on **C**, $R^2=0.93$) and horizontal velocity ($v_{xy}$) in the thermal (**D**, $R^2=0.97$). (See also Figure S3, Table S2 and S4.)

## Cross-country Optimization Strategy

When birds fly long distances (sometimes referred to as cross-country flight), they rely on multiple thermals and glide between them. Birds may have strategies where thermalling and gliding flight are "linked" to achieve a specific optimized goal. For example, these strategies could maximize the overall distance travelled for a given period, which takes into account the strengths of the thermals and aerodynamic constraints that shape the polar curve. To investigate cross-country optimization strategies between species, we explored how inter-thermal horizontal speed depends upon thermal strength. We used a linear approximation $v_{xy} = A\, v_{climb}^{Thermal} + B$ to represent the relationship between horizontal speed during inter-thermal flight, $v_{xy}$, and climb speed, $v_{climb}^{Thermal}$. The slope of the fitted line ($A$) represents *thermal-strength adaptivity*, i.e. how much the inter-thermal horizontal speed depends on thermal strength (Fig. 4A). The intercept ($B$) captures the *preferred inter-thermal gliding speed in zero thermal conditions*.

We calculated the thermal-strength adaptivity ($A_{Observed}$) by fitting the model to the observed flight data to evaluate the different species' behaviour and compare it to the optimum based on MacCready theory ($A_{MacCready}$). Importantly, while $A_{Observed}$ only depended on the observed data without explicit assumptions about their polar curve, $A_{MacCready}$ could only be calculated with the assumption that this polar curve was known. Here we used our estimated empirical polar curves to estimate the speeds suggested by the MacCready formula.

Based on this comparison, we found that the behaviour of the species could be roughly divided into three different categories (1) $A_{Observed} > 0$ and $A_{Observed} \sim A_{MacCready}$, (2) $A_{Observed} > 0$ and $A_{Observed} < A_{MacCready}$, (3) $A_{Observed} \sim 0$ (See Fig. 4B-D), where we mean $A_{Observed} > 0$ when it is significantly larger than 0 as compared to the randomization test and $A_{Observed} \sim 0$ otherwise (Tables S5, see more details below).

We found large variations in $A_{Observed}$ across species, meaning that species adopt different strategies when choosing their inter-thermal horizontal speed as a function of the strength of the thermals. Six species (*F. naumanni, F. peregrinus, A. rapax, A. nipalensis, G. himalayensis,* and adult *G. fulvus*) fall within group 1, as their $A_{Observed}$ was relatively high (i.e. significantly larger than 0 compared by randomization test, Table S5). This means their gliding speeds depended strongly on thermal strength. In addition, their cross-country optimization reached close to the full potential suggested by the MacCready theory, as shown by their $A_{Observed}$ close to $A_{MacCready}$ (Fig 4A, taking into account the accuracy of the parameter estimation). *F. naumanni* applied the highest degree of optimization among the species ($A_{Observed}$ = 4.5). *A. rapax* had the second highest slope ($A_{Observed}$ = 3.2). Except for both eagles (*A.rapax* and *A.nipalensis*), species chose similar, although somewhat lower, *B* values to the MacCready optimum (Fig. 4C), which means taking a bit slower horizontal speed between thermals.

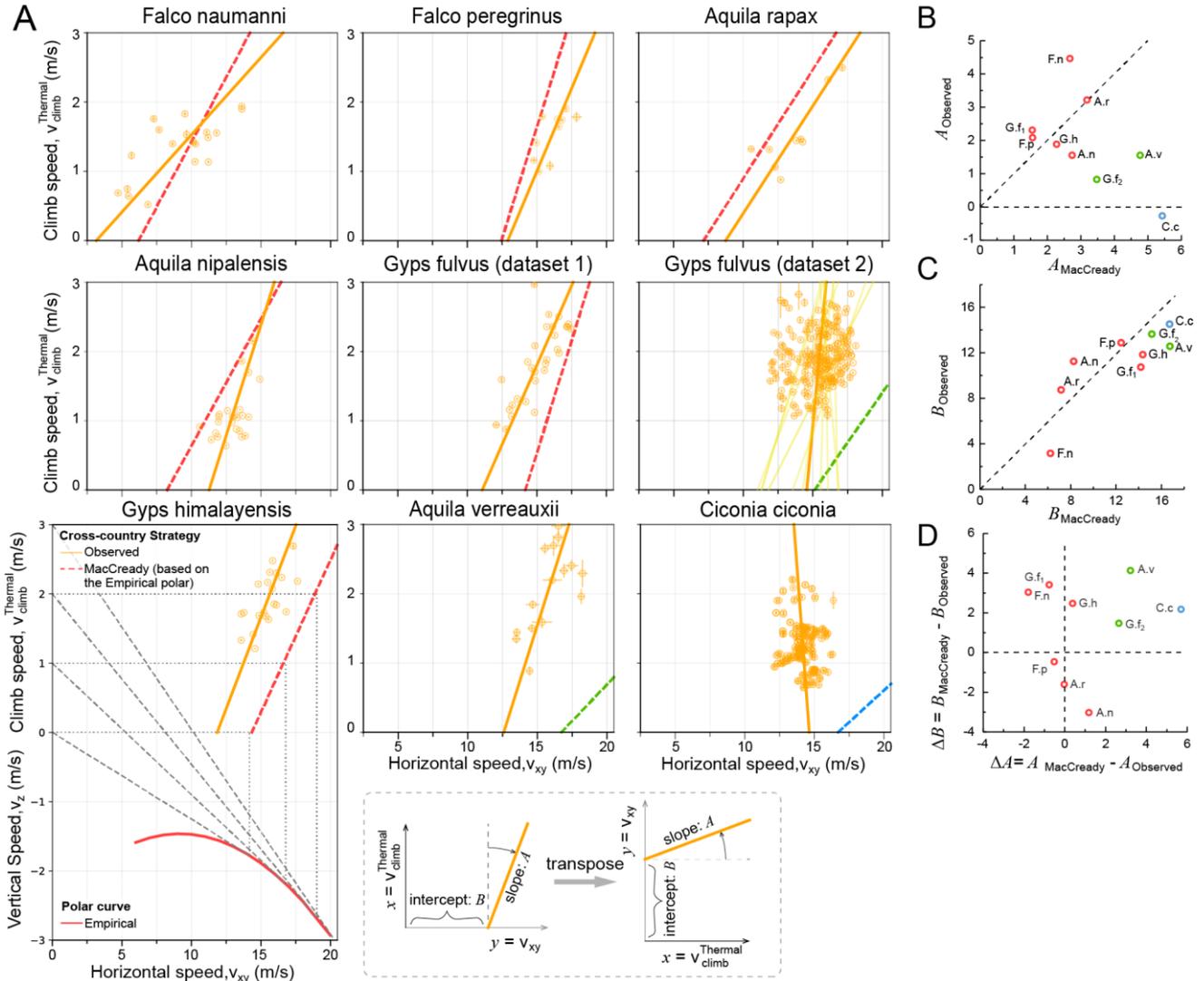

**Figure 4: Evaluating cross-country optimization strategies for different species. A)** Relationships between climb speed and inter-thermal horizontal speed for 8 species (species with at least 10 different flight days, see Methods). Large plot at bottom left provides full overview of the analysis using data of *G. himalayensis*. Solid (red) line shows empirical polar curve at bottom half. Dashed (red) line at top half represents the optimal soaring strategy using MacCready theory. Orange circles show daily individual means, the orange lines were fitted to the data. Note that the climb speed (y-axis) is the independent variable used on the linear fit (slope *A* represents *thermal-strength adaptivity,* and intercept (*B*) captures the *preferred inter-thermal gliding speed in zero thermal conditions*; see inset). For the other species, empirical polar curves are not shown. The colours indicate the three observed cross-country optimization strategies: red - thermal strength-dependent strategy according to MacCready, green - thermal strength-dependent optimisation sub-optimal to MacCready, blue - choosing gliding speed independent of thermal strength. We used a linear approximation of the optimal soaring curve, for easier comparison to the real data. **B-C)** Scatter plots showing *A* (and *B*) values for the species from the line fitted to flight data ($A_{Observed}$ and $B_{Observed}$, respectively) versus the value ($A_{MacCready}$ and $B_{MacCready}$, respectively) of optimal gliding strategy as calculated using the MacCready theory from the empirical polar fitted to the gliding data. **D)** Scatter plot indicating differences between the observed and the MacCready suggested parameters for *A* and *B*. (See also Figures S4 and S5.)

Vultures also seemed to be able to adjust their flight speed according to their daily climb speed, and the optimization tendency of *G. himalayensis* ($A_{\text{Observed}}$ = 1.89, p = 0.002, Table S5) was similar to adult *G. fulvus* of dataset 1 ($A_{\text{Observed}}$ = 2.32, p < 0.001, Table S5), but not for the mixed-aged birds of dataset 2.

In contrast, two species (*A. verreauxii* and the mixed-aged dataset of *G. fulvus*) can be categorized as group 2 where birds apply a thermal strength adaptive strategy ($A_{\text{Observed}}$>0) when choosing their inter-thermal speed, but the level of adaptivity is considerably smaller than suggested by the MacCready theory ($A_{\text{Observed}} < A_{\text{MacCready}}$). We analysed the two *G. fulvus* datasets separately (Fig. S4), and found that the thermal strength adaptivity was much higher for dataset 1 ($A^{(1)}_{\text{Observed}} > A^{(2)}_{\text{Observed}}$). Thus, birds of dataset 2 used suboptimal speeds compared to MacCready while birds of dataset 1 used it close to optimal ($A^{(1)}_{\text{Observed}} \sim A^{(1)}_{\text{MacCready}}$). We investigated the differences in detail later. For *A. verreauxii* respective $B_{\text{Observed}}$ values were also much smaller as compared to the optimal ($B_{\text{Observed}} < B_{\text{MacCready}}$),

The third group contained only a single species from our datasets, C. *ciconia*. They did not apply an adaptive strategy for choosing their cross-country speed based on thermal strength, as their $A_{\text{Observed}}$ was close to zero. The data came from multiple individuals that fly as a flock, so this could be an effect of a collective decision on the flight speed selection [34].

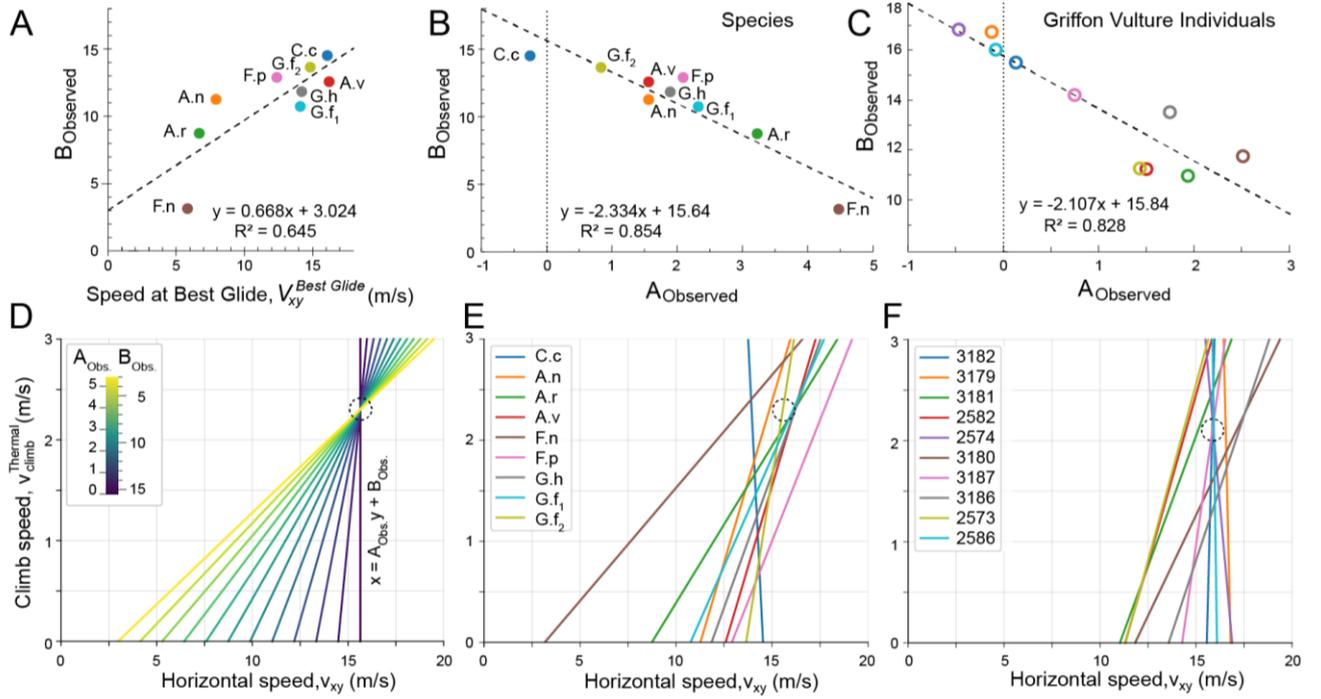

**Figure 5: Relationship between the parameters of cross-country optimization strategy for all species and for individual Griffon vultures.** To characterize cross-country behaviour, we fitted for each species a line to the horizontal speed $v_{xy}$ during inter-thermal flight as a function of the climb speed $v_{climb}^{Thermal}$ (that relates to the strength of the thermals) as $v_{xy} = A_{Obs.} \cdot v_{climb}^{Thermal} + B_{Obs.}$. **A)** Relationship between $B_{Observed}$ and the horizontal speed at best glide. **B)** Relationship between $B_{Observed}$ and $A_{Observed}$ for all species. Dashed line shows the linear fit to the data. **C)** As B, just for individual Griffon vultures, as dataset 2 was large enough to analyse the cross-country behaviour individually. **D)** Plotting possible lines on the $v_{climb}^{Thermal}$ and $v_{xy}$ diagram (as in Fig. 4A) for different $A_{Obs.}$ and $B_{Obs.}$ values indicated by their colour coding, ranging $A_{Obs.} = 0$ (no thermal adaptivity, dark blue) to $A_{Obs.} = 5$ (a highly adaptive strategy, yellow). The relationship between $A_{Obs.}$ and $B_{Obs.}$ means that they are coupled, and thus the lines follow a pattern where each goes through a single point (marked with the dashed circle). The coordinates of this point are determined by the coefficients of the linear fit shown on panel B, (x=15.64, y=2.334). Note that the inter-thermal horizontal speed, $v_{xy}$ (x-axis) is a function of the climb speed, $v_{climb}^{Thermal}$ (y-axis). **E-F)** The cross-country strategy lines are shown for each species (on **E**) and for individual Griffon vulture (**F**), using the same axes and ranges as on **D**. The dashed circles show the point defined by the fitted line on panels **B** and **C**, respectively. The colour codes match the respective plots on **B** and **C**. (See also Table S5.)

Finally, we explored whether there is a general relationship between the parameters $A_{Observed}$ and $B_{Observed}$ across the species to understand which thermal-strength dependent strategy birds use. The adaptivity ($A_{Observed}$) was strongly correlated to the preferred no-thermal gliding speed ($B_{Observed}$), meaning that the observed cross-country strategy birds apply has only one free parameter (instead of two; negative correlation, $R^2 = 0.85$, n=9, p = 0.0001 Fig. 5B). This was also the case when examining individual griffon vultures ($R^2 = 0.83$, n=10, p = 0.0002, Fig. 5C). Birds with lower inter-thermal gliding speed in zero thermal conditions *(B*$_{Observed}$*)* were using a highly adaptive strategy (high values of *A*$_{Observed}$), and vice versa. Also, we found a strong correlation between *B*$_{Observed}$ and the horizontal speed at best glide ($R^2 = 0.65$, n=9, $p = 0.0076$, Fig. 5A). Thus, birds seem to optimise their flight speed by typically using the "best glide" when thermals are weak. Overall, the linear trend found between *A*$_{Observed}$ and *B*$_{Observed}$ (Fig. 5B) defined a relationship that was general throughout all studied species (and individuals within species, Fig. 5C), so represents a characterisation of the cross-country strategy (Fig. 5D for the idealistic linear relationship, then for real data on Fig. 5E-F).

## Discussion

A comprehensive tracking data set of birds freely flying has enabled us to quantify how morphology and thermal conditions affect flight performance and behaviour in different soaring species. The growing literature studying flight behaviour of soaring birds by using high-frequency GPS data [35–39] allowed us to do a systematic comparison, to assess previous theoretical predictions and to discover a general rule describing cross-country optimisation.

During flight, soaring birds exploit ascending currents to travel large distances without or only a reduced amount of flapping flight. Our results confirmed the aerodynamic theory, as we show relationships between the wing loading and horizontal speed at crucial points of the polar curve (i.e. maximum glide ratio and minimum sink speed). Furthermore, we found that species whose wing loading was higher typically flew faster, and consequently turned on a larger radius, than lighter ones (Fig. 3C). The combination of circling radius and minimum sink speed determines the maximum benefits soaring birds can obtain from thermals. Since both flight aspects primarily depend on wing loading, species with lower wing loading outperform those with higher wing loading in the same thermal.

These species-specific features and choices contribute to the observed general cross-country trend. We explored how different species adapt their cross-country strategies depending on thermal strength (Fig. 4). We found that the preferred no-thermal gliding speed $B_{Observed}$ was close to the maximum glide ratio (that allows the birds to travel the farthest from a given altitude), meaning that birds optimise gliding speed in relation to their aerodynamic properties (Fig. 5A). We observed a negative relationship between $A_{Observed}$ and $B_{Observed}$ across all species (and between individuals for a single species). We also show that this thermal strength-dependent optimization strategy was highly related to certain morphological parameters. More specifically, species (individuals) with lower wing loading adopt a strategy where their inter-thermal gliding speed depends more strongly on thermal strength. Lower wing loading allows them to circle closer to the core of the thermal (where it is the strongest) and experience stronger lifts [33]. Also, our results show that the average horizontal speed in the thermals was correlated to the horizontal speed at minimum sink speed ($R^2 = 0.73$), which allows a bird to take advantage of even weak thermals. We found that wing loading was a major deterministic morphological feature that defines the horizontal speed at the minimum sink speed (as seen in Fig. 3B).

The combination of these two effects (i.e. the relationships of wing loading to inter-thermal glide speed and minimum sink speed) allows birds with lower wing loading to gain even higher benefits in the thermal, reaching higher climb speeds. On the other hand, these birds can afford to choose their inter-thermal speed more boldly and thus travel at a higher speed in good thermalling conditions. Our thermal-strength adaptivity ($A_{Observed}$) results depend not on the glide polar and the related MacCready speed-to-flight theory [24], but only on horizontal speed selection related to the daily average climb speed. MacCready theory does not take into account other environmental parameters for cross-country flights, such as the number of exploited thermals per day or the distance between thermals. Following MacCready theory causes a risk of being grounded or to avoid that, the need for switching to costly flapping flight for birds [18]. Other environmental factors[40,41] could cause similar risks. For this reason, birds that show a high degree of thermal-strength adaptivity can be considered more risk-prone species.

Furthermore, our results confirm the previous findings about *F. peregrinus* and *A. nipalensis* that the relation between flight speed and updraft strength is correlated and follows MacCready theory[25,26,42].

Variation between bird species in morphological features and flight style was related to the species' behaviour and ecological needs [11,43]. In general, birds' wing shapes are generally evolved to minimize the energy costs of flying at their typical speed and flight mode [43,44] but they still need to perform other phases of flight, such as take-off and aerial attack, which are related to their species-specific niches. For our species, although all raptors rely heavily on thermals to facilitate low-cost foraging flight, *F. peregrinus* requires more agility to capture aerial prey. In contrast, eagle species employ powerful attacks on prey on the ground and in the air. Similarly, all scavengers need to maximize their flight distance to locate carcasses, and a high capability for manoeuvrability is not necessary, as they do not typically attack moving animals. *C. ciconia* and *G. eremita* predominantly feed on the ground and perform long-range migration. Thus, adapting to these species-specific environments results in variations in the flight performance and behaviour of bird species. For this reason, maximizing the flight range might not be the only purpose during the flight. There might be several different optimizations by birds, such as keeping the prey attack range, maximizing the flight duration or migrating as a flock.

Apart from species differences, individual birds may also exhibit variation in flight performance due to differences in lifetime stage (e.g. age, breeding status, migration strategy). For example, older soaring birds may tolerate larger amounts of flapping flight during migration to reach their breeding grounds earlier [45]. Alternatively, birds with different experiences may differ in their skills to exploit thermals, or soar under challenging wind conditions [28] Here we observed notable inter-individual differences in the strategies of Griffon vulture individuals. Adult griffon vultures vary in their flight speeds and flight height according to the motivation and flight purpose (migration vs foraging flight, outbound vs inbound flight[28]). Thus, experience or developmental stage (e.g. developments of flight muscles) may cause variations in thermal strength adaptivity and the related selection of horizontal speed in the mixed-aged data set of the Griffon vultures. It has been previously reported that adult vultures demonstrate superior abilities compared to juveniles in utilising thermals, even though they have similar wing loading [28]. A similar effect probably applies to inter-thermal speed selection, since experienced birds use tailwind more efficiently [28] and may assess the location of the next thermals better. On the other hand, birds that take advantage of the same thermal within a similar period can use visual cues from birds ahead that already discovered the thermal, allowing them to use the thermal core directly [46,47]. This use of social information increases the efficiency of their net altitude gain in thermals, which may directly lead to changes in their strategic choices. Moreover, during long-range cross-country flights, visual cues can help them predict the abundance and locations of thermals, which can significantly influence their strategy [48]. In summary, although there are various factors at individual- and species-level that can influence the flight performance of soaring birds, our comparative analyses discovered a general, empirical rule that describes the cross-country strategies across all analysed species.

# Methods

## *Datasets*

GPS-tracking data recorded on freely flying birds allowed us to quantify the effects of morphology on flight performance and behavioural strategies and a direct comparison among different species. We collected an extensive dataset from research groups around the world studying thermalling birds in the wild (previously published and unpublished data sets), and carried out a comparative analysis of 12 bird species: Lesser Kestrel (*Falco naumanni*)[38], Peregrine falcon (*Falco peregrinus*)[25], Bald Eagle (*Haliaeetus leucocephalus*), Verreaux's eagle (*Aquila verreauxii*)[23], Tawny eagle (*Aquila rapax*), Steppe Eagle (*Aquila nipalensis*)[37], Eurasian Griffon vulture (*Gyps fulvus*)[27,28], Himalayan vulture (*Gyps himalayensis*)[27], Rüppell's Vulture (*Gyps Rüppellii*), Andean condor (*Vultur gryphus*), White stork (*Ciconia ciconia*)[34,49,50], Northern bald ibis (*Geronticus eremita*)[21].

## *Data preparation*

Since the dataset was gathered from different devices and formats, we simplified the miscellaneous data in the same structure by taking only timestamp, longitude, latitude, and altitude recordings as the first step. The geodesic coordinates provided by the GPS were converted into metric coordinates using the locally flat approximation with an $(x, y) = (0,0)$ origin at the beginning of data belonging to each bird and day. These coordinates were smoothed by a Gaussian filter depending on the sampling rate of the GPS recordings for each flight day of each individual in our custom Python codes. Then we calculated the horizontal components of the velocity ($v_x, v_y$) and acceleration ($a_x, a_y$). High-frequency GPS tracks contain all movement of birds, from take-off to landing, and may include flapping flight. For this reason, thermalling and gliding flight parts of the track were automatically identified based on curvature ($\kappa = (v_x a_y - v_y a_x)/(v_x^2 + v_y^2)^{1.5}$) and vertical speed ($v_z$) parameters (gliding: $|\kappa| < 0.01$ m$^{-1}$ and $v_z < 0$ m/s, thermalling: $|\kappa| > 0.01$ m$^{-1}$ and $v_z > 0$ m/s). Soaring occurred when birds made consecutive turns and when the average vertical ground speed was positive.

Therefore, thermalling parts were identified as segments with positive vertical speeds and positive curvature. Since thermals drift with the wind, we removed the effects of wind effect (for details[25]) to obtain a more reliable estimate of circling radius. Thermalling parts lasting longer than ($t = 30$ seconds) were defined as flying in thermal, and local wind velocity (speed and direction) was calculated using both horizontal components of the bird's velocity in the thermal as described in Ákos et al [25].

## *Evaluation of gliding and thermalling phases of flight*

For each of these ascending phases, we determined circling radius, the mean instantaneous horizontal speed (circling velocity), the mean instantaneous vertical speed (climbing rate). Gliding is a type of flight when the birds fly forward and lose height (typically flying without flapping their wings, but even if they flap, they sink). So, we identified and extracted "effective" gliding parts using above mentioned thresholds for vertical speed and curvature. To estimate airspeed during gliding, we subtracted average daily wind velocity (speed and direction, see above) from ground speed during gliding. Hereafter, all horizontal speed represents the estimated airspeed. The empirical polar curve was fitted to the measured average sinking and horizontal velocities during gliding flight. We used a second-order approximation to capture the main characteristic of the curves and fitted $f(x) = ax^2 + bx + c$ for determining the empirical glide polar. We used these glide polars to identify the important gliding parameters of each species. The maximum glide ratio, (the largest distance travelled from a given height) was calculated by drawing a tangent from the origin, or when the polar curve is approximated with a quadratic formula given above by $\sqrt{c/a}$. The speed at minimum sink airspeed was calculated by converting the quadratic function to vertex form, $f(x) = a(x - h)^2 + k$. Since the parabola is negative, the vertex represents the maximum point. The x-coordinate (the airspeed at minimum sink speed, *h*) is found using *-b/2a*, and the minimum sink speed (y-coordinate, *k*) is found by substituting the *h* back into the original quadratic function. The airspeed at minimum sink ($v_{xy}^{Min\ Sink}$) denotes the horizontal speed at which the descent is minimum. When choosing a horizontal airspeed higher than $v_{xy}^{Min\,Sink}$, the rate of descent increases, but this increment is initially modest, causing the glide ratio – the ratio of forward speed to descent – to keep rising. The glide ratio reaches its peak at the best-glide airspeed ($v_{xy}^{Best\,Glide}$). Above this critical value, the glide ratio steadily declines.

## *Cross-country strategy*

Selection of inter-thermal gliding speed affects the overall cross-country speed which is the ratio between the distance travelled during gliding divided by the total time (that includes both gliding and thermalling). By optimally selecting the inter-thermal gliding speed birds (and any aircrafts) can maximize the distance covered within the same amount of time, which is an essential factor during migration for migratory birds or during long-ranging foraging flights (searching for prey or carrion efficiently) for birds of prey and scavengers. To study the effect of thermal strength on inter-thermal horizontal speed, we selected species from our data set, which had multiple days to present large enough range for a linear fit. We selected species that had at least 10 different daily flight trajectories (allowing data coming from multiple individuals).

We used a linear approximation to represent the relationship between horizontal speed during inter-thermal flight, $v_{xy}$, and climb speed, $v_{climb}^{Thermal}$ (that relates to the strength of the thermals). Before fitting a line, we removed the outliers using Gaussian distribution and cut-off from two standard deviations. We estimated the observed strategy by fitting a line (*f(x) = Ax + B*) using the horizontal glide airspeed ($v_{xy}$) as *f(x)* that is the result of the bird's decision-making, and the climb speed ($v_{climb}^{Thermal}$) as *x* that is the input variable for the optimization. Here, parameter $A_{Observed}$ shows how adaptively the birds tune their inter-thermal speed to the climb speed, and a fit with $A_{Observed} = 0$ would indicate that the birds use the same inter-thermal speed irrespective of thermal strength. Birds with high values of $A_{Observed}$ fly between thermals much faster on days with strong thermals as opposed to weak thermals, while birds with $A_{Observed}$ close to zero do not vary their inter-thermal speed based on thermal strength. The intercept of that line, $B_{Observed}$, captures the preferred gliding speed in zero thermal conditions, representing the lowest value for daily average horizontal speed during gliding flight, thus we name it as *preferred no-thermal gliding speed* (which is the horizontal component).

## *Generating a theoretical glide polar using the Pennycuick's flight tool*

We used Pennycuick's Flight tool version 1.25 [12] to generate theoretical glide polars based on the morphological parameters. We set environment and aerodynamics parameters to remain consistent across all species to demonstrate morphology-related differences. All glide polars were created in the environmental conditions at an altitude of 1400 meters (most commonly used value at flight 1.25), and at the respective air density, 1.069 kg/m$^2$. The wingspan reduction law was configured to 'minimize induced + profile drag,' effectively flattening the glide polar at higher speeds. We initially generated the glide polars using the default settings of the software, which only differed in the induced drag factor (0.9). Later, we employed the value for the same parameter found in the Modelling Flying Bird (1.1). In order to obtain more accurate estimations for the glide polars, we conducted literature research on body and wing profile drag coefficients from experimental studies [29,30,51,52]. From the newest wind tunnel research [30], we found a value of 0.25 for the body drag coefficient and 0.025 for the wing profile drag coefficient by averaging the given range for the swift (*Apus apus*). These updates on the drag coefficients allowed us to create closer glide polars to our empirical glide polars.

## *Phylogenetic Analysis*

In order to make a phylogenetic comparison, we collected the partial or complete coding DNA sequence (CDS) of mitochondrial cytochrome b of relevant species using https://www.ncbi.nlm.nih.gov/ database [53–58]. The genetic distance (Fig. S5) between these species was calculated using Mega 11 software with the pairwise p-distance model. In this model, we used the gamma distribution with a shape parameter of 1 and employed pairwise deletion for handling missing data [59].

## *Randomization Test*

We applied a randomization test, which presents a methodology for evaluating the significance of linear relationships between variables. This approach entails maintaining the independent variable (x) constant while permuting the dependent variable (y) numerous times. By fitting a linear model and assessing the goodness of fit, typically measured through R-squared ($R^2$) metrics in our results, for the couples in each iteration, an empirical null distribution was built. Consequently, this enabled the determination of the extent to which the observed association between x and y deviates from what would be expected under random chance alone, yielding a reliable p-value for the significance of the linear relationship.

**Declaration of interests**

The authors declare no competing interests.

**Author Contributions**

G.K., O.D., and M.N. conceived the idea and designed the project;

O.D. collected experimental data; G.K. collected and standardized our original and the previously published data sets;

G.K. and M.N. designed the data analysis with contribution from O.D. and A.F;

G.K., P.L. and M.N. analysed the data;

G.K., A.F, and M.N. wrote the paper with contributions from O.D. and P.L.;

All authors revised the manuscript.

**Acknowledgement**

We thank authors who shared their published data, namely Roi Harel, Jesus Hernandez-Pliego, Megan Murgatroyd, Ran Nathan, Kate Reynolds, Graham Taylor, Bernhard Voelkl and Johannes Fritz. G.K was supported by Stipendium Hungaricum. This research was partially supported by Eötvös Loránd University and the Hungarian Academy of Sciences (grant number 95152). This project was partially supported by the National Research, Development and Innovation Office under grant no. K128780.

M.N. acknowledges support from the Hungarian Academy of Sciences, Grant 95152 (to the MTA-ELTE "Lendület" Collective Behaviour Research Group), the National Research, Development and Innovation Office under grant no. K128780, and the Isaac Newton Institute for Mathematical Sciences for support and hospitality during the programme 'Mathematics of Movement: an interdisciplinary approach to mutual challenges in animal ecology and cell biology', supported by the EPSRC Grant Number EP/R014604/1.

P.L. was supported by Eötvös Loránd University, CollMot Robotics Ltd. and the Ministry of Culture and Innovation of Hungary from the National Research, Development and Innovation Fund (awarded to P.L., Project no. C1794246, KDP-2021 funding scheme).

O.D. acknowledges the staff of Rocher des Aigles, Rocamadour, France, for long-term support of experiments using captive raptors since 2010: R. Arnaud, D. Maylin, B. Nouzière, and all falconry staff. O.D. also thanks C. Tromp, Y. Ropert-Coudert, A. Kato and several students for data collection between 2010 and 2014. O.D. thanks Giacomo Dell'Omo and the team from TechnoSmart who provided high-frequency GPS tags and accelerometers.

AF was supported by the Germany Research Foundation (DFG, Emmy Noether Fellowship 463925853), the Max Planck Society, the Hans und Helga Maus-Stiftung, and the James Heinemann research award of the Minerva Stiftung.24

# References

1. Tobias, J.A., Sheard, C., Pigot, A.L., Devenish, A.J.M., Yang, J., Sayol, F., Neate-Clegg, M.H.C., Alioravainen, N., Weeks, T.L., Barber, R.A., et al. (2022). AVONET: morphological, ecological and geographical data for all birds. Ecol. Lett. *25*, 581–597. https://doi.org/10.1111/ele.13898.

2. Hedenström, A. (1997). Predicted and observed migration speed in Lesser Spotted Eagle Aquila pomarina. Ardea *85*, 29–35.

3. Shamoun-Baranes, J., Bouten, W., Van Loon, E.E., Meijer, C., and Camphuysen, C.J. (2016). Flap or soar? How a flight generalist responds to its aerial environment. Philos. Trans. R. Soc. B Biol. Sci. *371*, 20150395.

4. Lasiewski, R.C., and Dawson, W.R. (1967). A re-examination of the relation between standard metabolic rate and body weight in birds. Condor *69*, 13–23.

5. Pennycuick, C.J. (1972). Soaring behaviour and performance of some East African birds, observed from a motor-glider. Ibis (Lond. 1859). *114*, 178–218.

6. Askew, G.., and Ellerby, D.. (2007). The mechanical power requirements of avian flight. Biol. Lett. *3*, 445–448. https://doi.org/10.1098/rsbl.2007.0182.

7. Mueller, H.C. (1991). Flight Strategies of Migrating Hawks at JSTOR.

8. Spaar, R., and Bruderer, B. (1996). Soaring migration of Steppe Eagles Aquila nipalensis in southern Israel: flight behaviour under various wind and thermal conditions. J. Avian Biol., 289–301.

9. Pennycuick, C.J. (1998). Field observations of thermals and thermal streets, and the theory of

cross-country soaring flight. J. Avian Biol., 33–43.

10. Pennycuick, C.J. (2003). The concept of energy height in animal locomotion: separating mechanics from physiology. J. Theor. Biol. *224*, 189–203.

11. Norberg, U.M. (1990). Vertebrate Flight (Oxford University Press) https://doi.org/10.1002/j.2326-1951.1962.tb00477.x.

12. Pennycuick, C.J. (2008). Modelling the flying bird (Elsevier).

13. Norberg, U.M., and Rayner, J.M. V (1987). Ecological morphology and flight in bats (Mammalia; Chiroptera): wing adaptations, flight performance, foraging strategy and echolocation. Philos. Trans. R. Soc. London. B, Biol. Sci. *316*, 335–427.

14. Thomas, F., and Milgram, J. (1999). Fundamentals of Sailplane Design.

15. Shamoun-Baranes, J., Leshem, Y., Yom-Tov, Y., and Liechti, O. (2003). Differential use of thermal convection by soaring birds over central Israel. Condor *105*, 208–218.

16. Leshem, Y., and Yom-Tov, Y. (1996). The use of thermals by soaring migrants. Ibis (Lond. 1859). *138*, 667–674.

17. Spaar, R. (1997). Flight strategies of migrating raptors; a comparative study of interspecific variation in flight characteristics. Ibis (Lond. 1859). *139*, 523–535.

18. Horvitz, N., Sapir, N., Liechti, F., Avissar, R., Mahrer, I., and Nathan, R. (2014). The gliding speed of migrating birds: Slow and safe or fast and risky? Ecol. Lett. *17*, 670–679. https://doi.org/10.1111/ele.12268.

19. Shannon, H.D., Young, G.S., Yates, M.A., Fuller, M.R., and Seegar, W.S. (2002). American white pelican soaring flight times and altitudes relative to changes in thermal depth and

intensity. Condor *104*, 679–683.

20. Weimerskirch, H., Le Corre, M., Ropert-Coudert, Y., Kato, A., and Marsac, F. (2005). The three-dimensional flight of red-footed boobies: adaptations to foraging in a tropical environment? Proc. R. Soc. B Biol. Sci. *272*, 53–61.

21. Voelkl, B., and Fritz, J. (2017). Relation between travel strategy and social organization of migrating birds with special consideration of formation flight in the northern bald ibis. Phil. Trans. R. Soc. B *372*.

22. Hernández-Pliego, J., Rodríguez, C., Dell'Omo, G., and Bustamante, J. (2017). Combined use of tri-Axial accelerometers and GPS reveals the flexible foraging strategy of a bird in relation to weather conditions. PLoS One *12*, 1–29. https://doi.org/10.1371/journal.pone.0177892.

23. Murgatroyd, M., Photopoulou, T., Underhill, L.G., Bouten, W., and Amar, A. (2018). Where eagles soar: Fine-resolution tracking reveals the spatiotemporal use of differential soaring modes in a large raptor. Ecol. Evol. *8*, 6788–6799. https://doi.org/10.1002/ece3.4189.

24. MacCready, P.B. (1958). Optimum airspeed selector. Soaring *11*, 10.

25. Akos, Z., Nagy, M., and Vicsek, T. (2008). Comparing bird and human soaring strategies. Proc. Natl. Acad. Sci. U. S. A. *105*, 4139–4143. https://doi.org/10.1073/pnas.0707711105.

26. Taylor, G.K., Reynolds, K. V., and Thomas, A.L.R. (2016). Soaring energetics and glide performance in a moving atmosphere. Philos. Trans. R. Soc. B Biol. Sci. *371*. https://doi.org/10.1098/rstb.2015.0398.

27. Duriez, O., Kato, A., Tromp, C., Dell'Omo, G., Vyssotski, A.L., Sarrazin, F., and Ropert-Coudert, Y. (2014). How cheap is soaring flight in raptors? A preliminary investigation in freely-flying vultures. PLoS One *9*. https://doi.org/10.1371/journal.pone.0084887.

28. Harel, R., Horvitz, N., and Nathan, R. (2016). Adult vultures outperform juveniles in challenging thermal soaring conditions. Sci. Rep. *6*, 1–8. https://doi.org/10.1038/srep27865.

29. Lentink, D., Müller, U.K., Stamhuis, E.J., De Kat, R., Van Gestel, W., Veldhuis, L.L.M., Henningsson, P., Hedenström, A., Videler, J.J., and Van Leeuwen, J.L. (2007). How swifts control their glide performance with morphing wings. Nature *446*, 1082–1085. https://doi.org/10.1038/nature05733.

30. Henningsson, P., and Hedenström, A. (2011). Aerodynamics of gliding flight in common swifts. J. Exp. Biol. *214*, 382–393.

31. Anderson, J.D. (1999). Aircraft performance and design. (WCB/McGraw-Hill).

32. Ákos, Z., Nagy, M., Leven, S., and Vicsek, T. (2010). Thermal soaring flight of birds and unmanned aerial vehicles. Bioinspiration and Biomimetics *5*. https://doi.org/10.1088/1748-3182/5/4/045003.

33. Williams, H.J., Duriez, O., Holton, M.D., Dell'Omo, G., Wilson, R.P., and Shepard, E.L.C. (2018). Vultures respond to challenges of near-ground thermal soaring by varying bank angle. J. Exp. Biol. *221*. https://doi.org/10.1242/jeb.174995.

34. Flack, A., Nagy, M., Fiedler, W., Couzin, I.D., and Wikelski, M. (2018). From local collective behavior to global migratory patterns in white storks. Science (80-. ). *360*, 911–914. https://doi.org/10.1126/science.aap7781.

35. Shepard, E.L.C., Lambertucci, S.A., Vallmitjana, D., and Wilson, R.P. (2011). Energy beyond food: foraging theory informs time spent in thermals by a large soaring bird. PLoS One *6*, e27375.

36. Bohrer, G., Brandes, D., Mandel, J.T., Bildstein, K.L., Miller, T.A., Lanzone, M., Katzner, T.,

Maisonneuve, C., and Tremblay, J.A. (2012). Estimating updraft velocity components over large spatial scales: contrasting migration strategies of golden eagles and turkey vultures. Ecol. Lett. *15*, 96–103. https://doi.org/10.1111/j.1461-0248.2011.01713.x.

37. Reynolds, K. V., Thomas, A.L.R., and Taylor, G.K. (2014). Wing tucks are a response to atmospheric turbulence in the soaring flight of the steppe eagle Aquila nipalensis. J. R. Soc. Interface *11*. https://doi.org/10.1098/rsif.2014.0645.

38. Hernández-Pliego, J., Rodríguez, C., and Bustamante, J. (2015). Why Do Kestrels Soar? PLoS One *10*. https://doi.org/10.1371/journal.pone.0145402.

39. Longarini, A., Duriez, O., Shepard, E., Safi, K., Wikelski, M., and Scacco, M. (2023). Effect of harness design for tag attachment on the flight performance of five soaring species. Mov. Ecol. *11*, 1–14. https://doi.org/10.1186/s40462-023-00408-y.

40. Shepard, E.L.C., Williamson, C., and Windsor, S.P. (2016). Fine-scale flight strategies of gulls in urban airflows indicate risk and reward in city living. Philos. Trans. R. Soc. B Biol. Sci. *371*, 20150394.

41. Richardson, P.L., Wakefield, E.D., and Phillips, R.A. (2018). Flight speed and performance of the wandering albatross with respect to wind. Mov. Ecol. *6*, 1–15.

42. Reynolds, K. (2015). Soaring and gust response in the Steppe Eagle.

43. Gill, F.B., and Prum, R.O. (2019). Ornithology (Macmillan).

44. Santos, C.D., Hanssen, F., Muñoz, A.R., Onrubia, A., Wikelski, M., May, R., and Silva, J.P. (2017). Match between soaring modes of black kites and the fine-scale distribution of updrafts. Sci. Rep. *7*, 1–10. https://doi.org/10.1038/s41598-017-05319-8.

45. Aikens, E.O., Nourani, E., Fiedler, W., Wikelski, M., and Flack, A. (2024). Learning shapes the development of migratory behavior. Proc. Natl. Acad. Sci. *121*. https://doi.org/10.1073/pnas.2306389121.

46. Nagy, M., Couzin, I.D., Fiedler, W., Wikelski, M., and Flack, A. (2018). Synchronization, coordination and collective sensing during thermalling flight of freely migrating white storks. Philos. Trans. R. Soc. B Biol. Sci. *373*. https://doi.org/10.1098/rstb.2017.0011.

47. Williams, H.J., King, A.J., Duriez, O., Börger, L., and Shepard, E.L.C. (2018). Social eavesdropping allows for a more risky gliding strategy by thermal-soaring birds. J. R. Soc. Interface *15*. https://doi.org/10.1098/rsif.2018.0578.

48. Sassi, Y., Nouzières, B., Scacco, M., Tremblay, Y., Duriez, O., and Robira, B. (2024). The use of social information in vulture flight decisions. Proc. R. Soc. B Biol. Sci. *291*. https://doi.org/10.1098/rspb.2023.1729.

49. Flack, A., Fiedler, W., and Wikelski, M. (2017). Data from: Wind estimation based on thermal soaring of birds.

50. Weinzierl, R., Bohrer, G., Kranstauber, B., Fiedler, W., Wikelski, M., and Flack, A. (2016). Wind estimation based on thermal soaring of birds. Ecol. Evol. *6*, 8706–8718.

51. Maybury, W.J. (2000). The aerodynamics of bird bodies.

52. Ortal, H. (2012). Turbulent Flow Characterization and Aerodynamic Forces Around a Gliding Osprey Model in a Wind Tunnel.

53. Griffiths, C.S. (1997). Correlation of Functional Domains and Rates of Nucleotide Substitution in Cytochromeb. Mol. Phylogenet. Evol. *7*, 352–365. https://doi.org/10.1006/mpev.1997.0404.

54. Groombridge, J.J., Jones, C.G., Bayes, M.K., Van Zyl, A.J., Carrillo, J., Nichols, R.A., and Bruford, M.W. (2002). A molecular phylogeny of African kestrels with reference to divergence across the Indian Ocean. Mol. Phylogenet. Evol. *25*, 267–277. https://doi.org/10.1016/S1055-7903(02)00254-3.

55. Lerner, H.R.L., and Mindell, D.P. (2005). Phylogeny of eagles, Old World vultures, and other Accipitridae based on nuclear and mitochondrial DNA. Mol. Phylogenet. Evol. *37*, 327–346. https://doi.org/10.1016/j.ympev.2005.04.010.

56. Arshad, M., Gonzalez, J., El-Sayed, A.A., Osborne, T., and Wink, M. (2009). Phylogeny and phylogeography of critically endangered Gyps species based on nuclear and mitochondrial markers. J. Ornithol. *150*, 419–430. https://doi.org/10.1007/s10336-008-0359-x.

57. Eder, H., Fiedler, W., and Neuhäuser, M. (2015). Evaluation of aerodynamic parameters from infrared laser tracking of free-gliding white storks. J. Ornithol. *156*, 667–677. https://doi.org/10.1007/s10336-015-1176-7.

58. Treutlein, T., and Wink, M. (2016). Phylogeny of birds inferred from the complete sequence of mitochondrial DNA. Phylogeny birds inferred from Complet. Seq. mitochondrial DNA. https://www.ncbi.nlm.nih.gov/nuccore/AY567909.1.

59. Tamura, K., Stecher, G., and Kumar, S. (2021). MEGA11: Molecular Evolutionary Genetics Analysis Version 11. Mol. Biol. Evol. *38*, 3022–3027. https://doi.org/10.1093/molbev/msab120.

# Supplementary Materials

for

## Adaptive cross-country optimisation strategies in thermal soaring birds

by

**Authors:** Göksel Keskin[1,2], Olivier Duriez[3], Pedro Lacerda[1,2], Andrea Flack[4], Máté Nagy[1,2,5]

**Affiliations:**

[1]MTA-ELTE Lendület Collective Behaviour Research Group, Hungarian Academy of Sciences, Budapest, Hungary
[2]Department of Biological Physics, Eötvös Loránd University, Budapest, Hungary
[3]CEFE, Univ Montpellier, CNRS, EPHE, IRD, Montpellier, France
[4]Collective Migration Group, Max-Planck Institute of Animal Behavior, Konstanz, Germany
[5]Max-Planck Institute of Animal Behavior, Konstanz, Germany

**This file contains:**
5 Supplementary Figures: S1-S5
5 Supplementary Tables: S1-S5
Supplementary Text

# Supplementary Materials

**Supplementary Figures:**

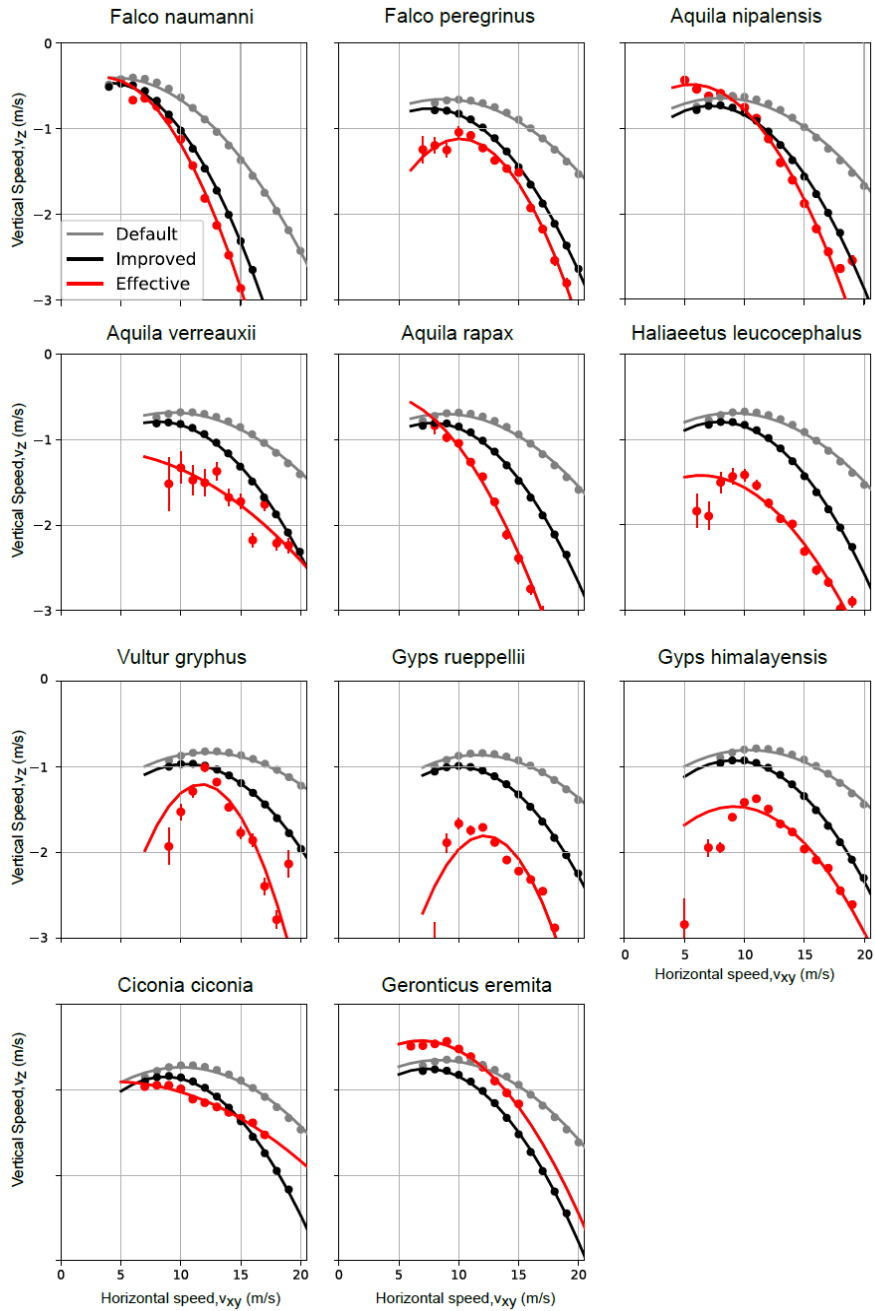

**Figure S1: Empirical polar curves and curves from Pennycuick estimations – related to Figure 2.** Three polar curves are presented, representing predicted glide performance for each species, except *Gyps fulvus*, presented in more detail in Fig. S4. Each data point on the curves indicates the corresponding **vertical speed (m/s)** (i.e. altitude loss) for a given **horizontal airspeed (m/s)**. These data points were used to fit the polar curves. The grey line shows the results using Pennycuick's Flight Tool v1.25 with default settings and the black line depicts the glide curve with parameter adjustments based on values from the literature. The red curve shows the glide polars created from observed gliding points in our dataset (see the methods for details).

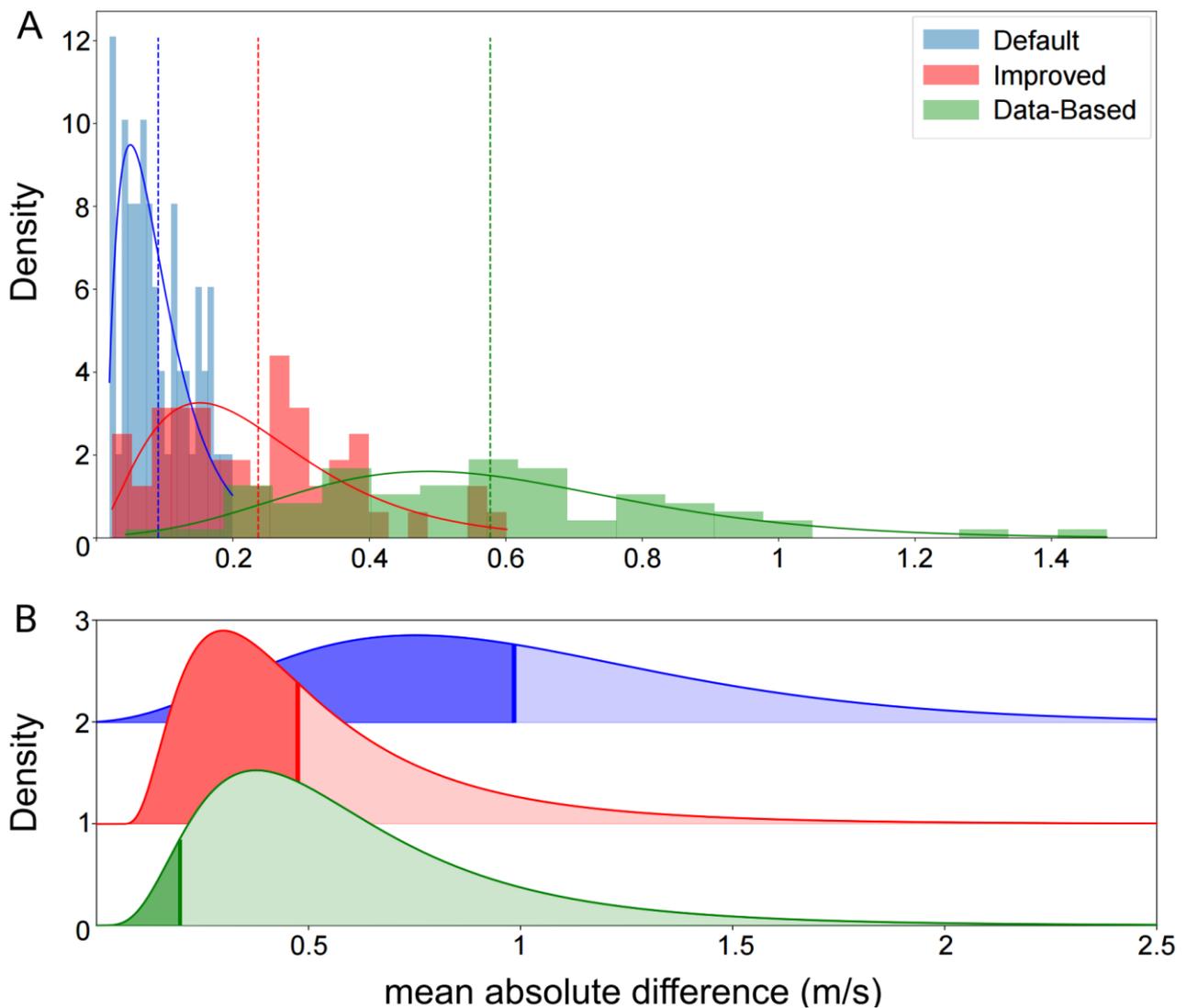

**Figure S2: Variation in glide polars created by Pennycuick's Flight tool and the empirical glide polar, and their significance to real data points with the randomization test – related to Figure 2. A)** The histogram plots show the distribution of the mean absolute difference between the different type of glide polars of species (Default: Glide polars created by the Pennycuick's flight tool using default settings, Improved: Glide polars created by using the Pennycuick's flight tool with an improved parameter set from the latest literature, see *Methods*). Smaller values indicate that the species-specific curves are very similar to each other, so they are not very tailored to any given species. Related colour curves to the histograms show the fitted function, while the vertical dashed lines indicate the mean value of the data. **B)** Mean absolute difference between the measured data points and polar curves for the three different types of polars ("Default", "Improved" and "Empirical", using the same colour as on panel A) shown with thick vertical lines. The distributions show the mean absolute distance values for randomization tests (see *Methods*), which were calculated between the measured data points and a randomly assigned glider polar from another species, to indicate the difference which is expected by chance. To assess significance, the p-value is calculated as the proportion of randomized test (10000 iterations) statistics that are less or equal to the observed value. This is indicated by the darker area on the left side of the distributions. Smaller values indicate better specificity to the species ($p<0.05$ only for the empirical glide polars. For the sake of legibility, the distributions were offset by one unit.

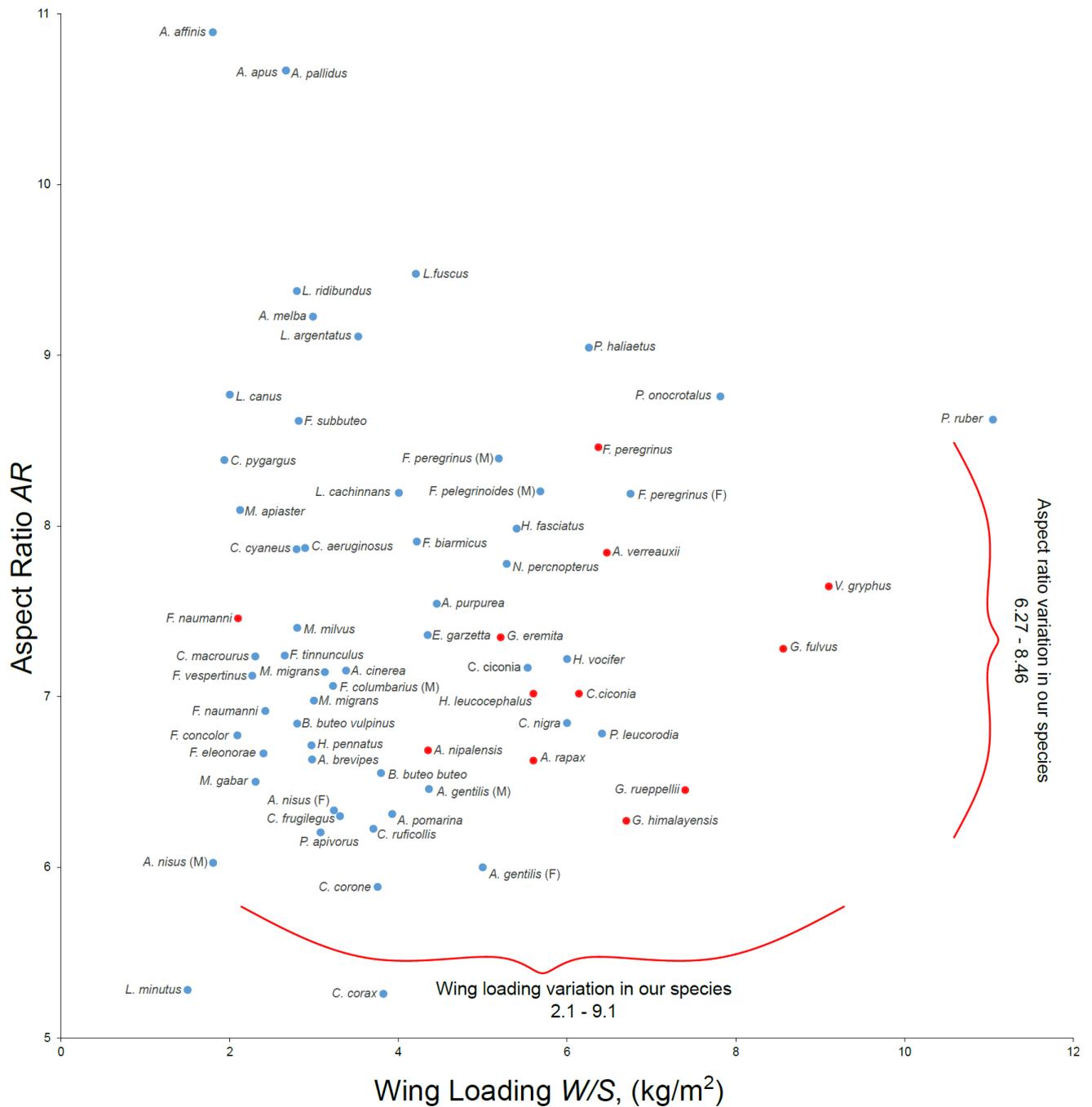

**Figure S3: Bird species exhibit widespread morphological differences, particularly in the intricate interplay between wing loading and aspect ratio – related to Figure 3.** Species marked in red in the figure were added by us, either measured or from the literature to a dataset previously created (Bruderer et al., 2010) with species known for gliding and soaring. The variation in our species is enough to assess the effect of wing loading (2.1 - 9.1) on flight performance and behaviour but not sufficient for the aspect ratio (6.27 - 8.46).

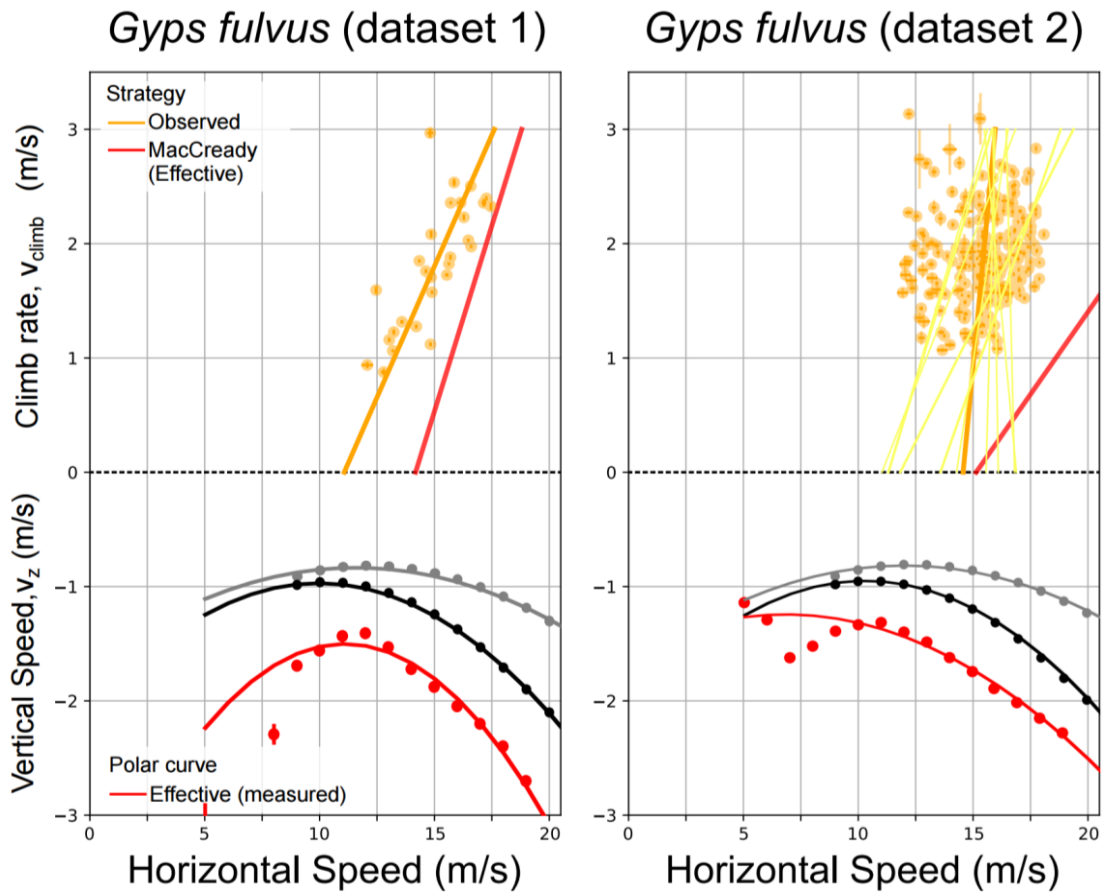

**Figure S4: Empirical (effective) polar curve and cross-country strategy for datasets 1 and 2 of Griffon Vultures – related to Figures 2 and 4.** Our data contains two different datasets belonging to Griffon Vultures, with the first dataset consisting of captive birds and the second one comprising wild birds, originating from France and around Israel respectively. As explained in Fig. 5, the bottom part of the figure shows the estimated glide polar (using Pennycuick flight tool default settings and improved values, please see methods and Fig. S1), empirical glide polar, and the soaring strategy curve positioned on the top. For dataset 2, we have sufficient data to show individual variation in soaring strategies (represented by yellow strategy curves). Notably, although the overall strategy line is different between the two datasets, there are individual birds from dataset 2 that closely match dataset 1.

| Optimization Strategy | Family | Capitive or Wild | | F.n | F.p | A.r | A.n | G.h | G.f$_1$ | G.f$_2$ | A.v | C.c |
|---|---|---|---|---|---|---|---|---|---|---|---|---|
| thermal strength-dependent, optimal according to MacCready theory | Falconidae | W | F.n | | **0.115** | 0.178 | 0.175 | 0.170 | 0.180 | 0.180 | 0.180 | 0.172 |
| | Falconidae | C | F.p | **0.115** | | 0.168 | 0.159 | 0.164 | 0.172 | 0.172 | 0.166 | 0.175 |
| | Accipitridae | C | A.r | 0.178 | 0.168 | | **0.029** | 0.120 | 0.123 | 0.123 | 0.065 | 0.163 |
| | Accipitridae | C | A.n | 0.175 | 0.159 | **0.029** | | 0.127 | 0.116 | 0.116 | 0.065 | 0.155 |
| | Accipitridae | C | G.h | 0.170 | 0.164 | 0.120 | 0.127 | | **0.033** | 0.033 | 0.116 | 0.154 |
| | Accipitridae | C | G.f$_1$ | 0.180 | 0.172 | 0.123 | 0.116 | **0.033** | | | 0.112 | 0.154 |
| thermal strength-dependent, sub-optimal | Accipitridae | W | G.f$_2$ | 0.180 | 0.172 | 0.123 | 0.116 | **0.033** | | | 0.112 | 0.154 |
| | Accipitridae | W | A.v | 0.180 | 0.166 | **0.065** | 0.065 | 0.116 | 0.112 | 0.112 | | 0.157 |
| independent of thermal strength | Ciconiidae | W | C.c | 0.172 | 0.175 | 0.163 | 0.155 | **0.154** | **0.154** | **0.154** | 0.157 | |

**Figure S5: Genetic closeness and cross-country strategy – related to Figure 4.** Additional information about the birds in our data set, grouped according to the level of optimisation discovered. The taxonomic family, whether birds were captive or wild, and the phylogenetic distances are given in a matrix format, based on genetic distance among the species. Colour coding indicates distance (green for the closest). For each line, the distance with the closest species is highlighted by bold, except for storks, which species is genetically the most distant from all the other species in this data set (see also Fig. 1).

**Supplementary Tables:**

**Table S1: Summary table showing the species, foraging style, the reference source, the number of individuals, the number of flight days and the gliding points that our data set contains high-resolution flight logs – related to Figure 1.** Foraging niche is based on (Pigot et al. 2020), except where it was unavailable (marked by *) we added based on personal observations.

| Species | Common Name | Foraging Niche | Data Source | Number of Birds | Individual Flight Days | Number of Gliding Points |
|---|---|---|---|---|---|---|
| *Falco naumanni* | Lesser Kestrel | Generalist, Invertivore air to surface (*) | Hernández-Pliego *et al.*, 2015[38] | 15 | 19 | 25,011 |
| *Falco peregrinus* | Peregrine Falcon | Vertivore aerial | Akos *et al.*, 2009[25] | 1 | 10 | 9,119 |
| *Haliaeetus leucocephalus* | Bald Eagle | Generalist air to surface (*) | Original (Previously unpublished) | 4 | 11 | 10,582 |
| *Aquila nipalensis* | Steppe Eagle | Vertivore air to surface | Reynolds *et al.*, 2014[37] | 1 | 24 | 81,460 |
| *Aquila verreauxii* | Verreauxii Eagle | Vertivore air to surface | Murgatroyd *et al.*, 2018[23] | 3 | 17 | 4,099 |
| *Aquila rapax* | Tawny Eagle | Vertivore air to surface | Original (Previously unpublished) | 2 | 9 | 5,575 |
| *Gyps fulvus* (dataset 1) | Griffon Vulture | Scavenger ground | Duriez *et al.*, 2014[27] | 4 | 30 | 85,900 |
| *Gyps fulvus* (dataset 2) | Griffon Vulture | Scavenger ground | Harel *et al.*, 2016[28] | 17 | 253 | 802,640 |
| *Gyps himalayensis* | Himalayan Vulture | Scavenger ground | Duriez *et al.*, 2014[27] | 2 | 21 | 107,805 |
| *Gyps rueppellii* | Rueppellii's Vulture | Scavenger ground | Original (Previously unpublished) | 1 | 6 | 19,054 |
| *Vultur gryphus* | Andean Condor | Scavenger ground | Original (Previously unpublished) | 1 | 7 | 3,164 |
| *Geronticus eremita* | Northern Bald Ibis | Invertivore ground | Voelkl *et al.*, 2017[21] | 11 | 11 | 97,284 |
| *Ciconia ciconia* | White Stork | Generalist ground (*) | Flack *et al.*, 2017[49,50] | 27 | 123 | 291,969 |

**Table S2: Morphological traits of species – related to Figure 3.** Data show values belonging to the individuals recorded if biometric measurements were available. Otherwise,

data from the literature were taken into account marking the source. (a: McGahan, 1973; b: Ferguson-Lees and Christie, 2010; c: Eder, Fiedler and Neuhäuser, 2015; d: Harel, Horvitz and Nathan, 2016; e: Mirzaeinia, Mirzaeinia and Hassanalian, 2020).

| Species | Wing Area, S (m$^2$) | Wingspan, b (m) | Aspect Ratio, AR (unitless) | Wing loading, W/S (kg/m$^2$) | Weight, W (kg) |
|---|---|---|---|---|---|
| *Falco naumanni* | 0.062 | 0.68 | 7.46 | 2.1 | 0.13 |
| *Falco peregrinus*[b] | 0.128 | 1.04 | 8.46 | 6.37 | 0.82 |
| *Haliaeetus leucocephalus* | 0.57 | 2 | 7.02 | 5.6 | 3.1 |
| *Aquila nipalensis* | 0.54 | 1.9 | 6.69 | 4.35 | 2.35 |
| *Aquila verreauxii* | 0.51 | 2 | 7.84 | 6.47 | 3.3 |
| *Aquila rapax* | 0.5 | 1.82 | 6.62 | 5.6 | 2.6 |
| *Gyps fulvus* (dataset 1) | 0.95 | 2.56 | 6.9 | 8.1 | 7.7 |
| *Gyps fulvus* (dataset 2)[d] | 0.95 | 2.7 | 7.6 | 9 | 8.5 |
| *Gyps himalayensis* | 1.25 | 2.8 | 6.27 | 6.7 | 8.0 |
| *Gyps rueppellii* | 0.75 | 2.2 | 6.45 | 7.4 | 5.6 |
| *Vultur gryphus*[a] | 1.1 | 2.9 | 7.65 | 9.1 | 10.0 |
| *Geronticus eremita*[e] | 0.23 | 1.3 | 7.35 | 5.21 | 1.2 |
| *Ciconia ciconia*[c] | 0.57 | 2 | 7.02 | 6.14 | 3.5 |

**Table S3: Parameters for the Pennycuick polar curves – related to Figure 2.** Default values of drag parameters in Pennycuick's Flight Tool 1.25 and drag parameters used to create improved glide polars and related literature are shown in the table. Note that all values are unitless.

|  | **Default** | **Improved** | **Main source for the improved parameters** | **Additional experimental studies** |
|---|---|---|---|---|
| Body drag coefficient | 0.1 | 0.25 | Swift, 0.22 - 0.30 (Henningsson and Hedenström, 2011) | Swift, 0.26 (Lentnik, 2009); Starling, 0.12 - 0.26, (Maybury, 2000); Harris Hawk, 0.18 (Tucker, 1990) |
| Induced drag factor | 0.9 | 1.1 | Swift, 1.1 (Henningsson and Hedenström, 2011) | Default in the book, 1.1 (Pennycuick, 2008) |
| Wing profile drag coefficient | 0.014 | 0.025 | Swift, 0.011-0.048 (Henningsson and Hedenström, 2011) | Osprey, 0.03 - 0.06 (Ortal, 2012); Harris Hawk, 0.008 - 0.052 (Pennycuick et al.,1992); Harris Hawk, 0.003 - 0.097 (Tucker and Heine, 1990) |

**Table S4 - Estimated flight metrics – related to Figure 3.** Important gliding and thermalling parameters derived from the empirical glide polar and thermalling part of the flight trajectories.

| Species | Best Glide Ratio | $v_{xy}$ at best glide | $v_z$ at best glide | $v_{xy}$ at min sink | $v_z$ at min sink | mean Radius | $v_{xy}$ Thermal |
|---|---|---|---|---|---|---|---|
| (Unit) | 1 | m/s | m/s | m/s | m/s | m | m/s |
| *Falco naumanni* | 11.8 | 5.81 | -0.49 | 3.57 | -0.4 | 8.92 | 4.7 |
| *Falco peregrinus* | 10.8 | 12.36 | -1.22 | 10.1 | -1.11 | 23.31 | 8.33 |
| *Haliaeetus leucocephalus* | 6.86 | 13.1 | -1.9 | 6.36 | -1.41 | 21.95 | 8.74 |
| *Aquila nipalensis* | 14 | 7.92 | -0.56 | 5.56 | -0.48 | 19.8 | 8.88 |
| *Aquila verreauxii* | 8.53 | 16.21 | -1.9 | 3.10 | -1.13 | 10.51 | 2.55 |
| *Aquila rapax* | 10.75 | 6.68 | -0.62 | 3.26 | -0.46 | 12.11 | 7.0 |
| *Gyps fulvus* (dataset 1) | 8.36 | 14.1 | -1.68 | 11.07 | -1.50 | 26.86 | 10.11 |
| *Gyps fulvus* (dataset 2) | 8.64 | 14.81 | -1.71 | 6.75 | -1.24 | 22.88 | 9.16 |
| *Gyps himalayensis* | 7.98 | 14.18 | -1.77 | 9.17 | -1.46 | 24.64 | 9.67 |
| *Gyps rueppellii* | 7.31 | 14.2 | -1.94 | 12.18 | -1.80 | 26.1 | 10.19 |
| *Vultur gryphus* | 10.31 | 13.07 | -1.26 | 11.7 | -1.20 | 21.5 | 9.6 |
| *Geronticus eremita* | 18.91 | 9.1 | -0.48 | 6.85 | -0.42 | 10.41 | 6.3 |
| *Ciconia ciconia* | 11.26 | 16.07 | -1.42 | 4.34 | -0.90 | 17.56 | 7.4 |

**Table S5: Thermal-strength adaptivity of species and comparison to randomization – related to Figure 5.** For each species where $A_{Observed}$ values were calculated we present the result of a randomization. For each species, the average daily climb speed and inter-thermal horizontal speed were permutated 1000 times, and the p gives the proportion that the $A_{rand}$ fitted to the permutated data set is larger than the real $A_{Observed}$. The mean and the standard deviation (SD) are given for the randomization. Significant p values are highlighted with bold text.

| Species | $A_{Observed}$ | p | Mean ($A_{rand}$) | SD($A_{rand}$) |
|---|---|---|---|---|
| *Aquila nipalensis* | 1.56 | **0.003** | 0.00 | 0.60 |
| *Aquila rapax* | 3.22 | **0.006** | -0.01 | 1.38 |
| *Aquila verreauxii* | 1.56 | **0.003** | -0.01 | 0.56 |
| *Ciconia ciconia* | -0.26 | 0.824 | 0.00 | 0.26 |
| *Falco peregrinus* | 2.09 | **0.011** | -0.03 | 1.01 |
| *Falco naumanni* | 4.48 | **<0.001** | -0.03 | 1.46 |
| *Gyps fulvus* (dataset 1) | 2.32 | **<0.001** | 0.01 | 0.54 |
| *Gyps fulvus* (dataset 2) | 0.83 | **<0.001** | 0.00 | 0.22 |
| *Gyps himalayensis* | 1.89 | **0.002** | 0.02 | 0.64 |

# Supplementary Text – Glossary

**Best glide**: the behaviour during which the glide ratio is maximal. Also indicated by a specific point on the polar curve.

**Glide ratio**: the distance travelled horizontally compared to the altitude lost

**Best glide ratio**: the largest glide ratio that ensures the maximum distance travelled horizontally for a given altitude lost

**Sink speed (rate)**: the vertical speed during altitude loss

**Minimum sink**: the behaviour during which the sink speed is minimal

**Minimum sink speed**: the vertical speed during minimum sink

**Climb rate**: the vertical speed during altitude gain

**Thermal**: localised atmospheric phenomenon caused by rising hot air

**Interthermal**: cruise flight between thermals

**Polar curve**: the functional relation between airspeed (horizontal speed) and the sink rate of an object (birds, aircraft)

**Aspect ratio**: the ratio of wing length to an average wing chord. This latter is typically calculated using ratio between the wing area and square of the wing length

**Wing Loading**: ratio of the body mass (weight) to the surface area of the wing. Depending on convention, it is typically given in units of $kg/m^2$ (using the mass) or $N/m^2$ (using the weight). In this paper we used the former definition.

**Gliding speed**: airspeed (horizontal speed) during gliding flight

**Cross-Country**: a long-distance flight behaviour using multiple thermals and gliding between them

**Cross-Country speed**: the average (horizontal) flight speed during cross-country flight

**Cross-Country strategy**: A specific strategy describing how Cross-Country speed is chosen. For example, depending on the strength of the thermals (the mean climb speed) or independent of that.